\documentstyle[sprocl]{article}

\def\Journal#1#2#3#4{{#1} {\bf #2}, #3 (#4)}

\def\PLB{{\em Phys. Lett.}  B}

\def\be{\begin{equation}}
\def\ee{\end{equation}}
\def\bea{\begin{eqnarray}}
\def\eea{\end{eqnarray}}
\def\J{$J/\psi$}

\begin{document}

\begin{flushright}
RIKEN--BNL Preprint\\
November 1998\\
hep-ph/9812214
\end{flushright}
\vskip1cm

\title{SUMMARY\footnote{Summary talk at the ``Quarkonium production in 
nuclear collisions" Workshop, May 11--15, 1998, INT, Seattle.}}

\author{D. KHARZEEV}

\vskip0.3cm

\address{RIKEN-BNL Research Center,\\ Brookhaven National Laboratory,
\\ Upton NY 11973, USA \\E-mail: kharzeev@bnl.gov} 

\vskip0.5cm

\maketitle\abstracts{
This summary is an attempt to overview  
the wealth of new results and ideas in quarkonium physics 
presented at the Seattle Workshop. }

\section{Introduction}

After the 1974 discovery of the $J/\psi$, the hadronic world was 
never the same. 
From the very beginning it had been recognized that this particle,   
with its unusually small width and large mass, was a stand-out in the 
hadronic zoo. The large mass was soon understood to be the consequence 
of the existence of a new massive quark flavor, and the small width -- 
the consequence of asymptotic freedom, making the coupling, 
and therefore, the annihilation probability, small at the scale of the 
charm quark mass. The understanding of $J/\psi$ properties   
was therefore crucial for establishing QCD as the standard model of 
strong interactions. 

Seattle Workshop took place at a particularly interesting time for 
quarkonium physics. The CERN SPS results from NA50 undoubtedly demonstrated 
that collective phenomena dramatically affect the \J\ production in Pb-Pb collisions. 
Fermilab Tevatron results of CDF and D0 made theorists to put under scrutiny the 
widely accepted color-singlet model of quarkonium production. 
Fermilab fixed target $pA$ results from E866 for the first time opened 
a new kinematic window of negative $x_F$. And, with RHIC experiments at BNL  
starting to accumulate data next year, we have to carefully analyze 
the meaning of the information available to us at present to make 
predictions for the future. 

What follows below is my personal account 
of the most significant results presented in Seattle; obviously, 
such a selection is always biased, and I could easily miss something 
very important. I apologize in advance also for potential  
misinterpretations of the presented talks -- to acquire a proper understanding, 
the reader is strongly encouraged to read the original contributions in this Volume.

\section{What have we learnt at the Workshop?}

The naive model of \J\ as a bound state of charm 
quark and anti-quark turned out to be only 
a rough approximation to reality, as {\bf E. Braaten}\cite{Eric} discussed 
at this Workshop, and one has to go beyond this model to 
explain the puzzles which exist not only in quarkonium production, 
but also in its decays. A particular problem addressed by Braaten at this Workshop 
was the strong violation of helicity selection rule observed in exclusive 
decays of \J\ into $\rho\pi$ final state. The helicity selection rule 
is a direct consequence of vector coupling of gluons to quarks, and 
should hold if all of the gluons involved in the process  
are hard. However, as was pointed out at by Braaten, 
the presence of dynamical gluons in the \J\ wave function leads 
to the possibility of substantial evasion of the helicity selection rule 
due to the $\bar{Q}Q$ annihilation in the color octet state. It is further 
argued that a similar evasion does not happen in the $\psi'$ decay as 
a consequence of the larger weight of the $\bar{D}D$ channel 
in the wave function, and the significant contribution of this 
channel depletes 
the role of the color-octet 
state at the distances $\sim 1/m_Q$ at which the annihilation takes place. 
It would be interesting to extend this analysis to other helicity--violating 
decays, e.g. $\eta_c \to \bar{p}p$. It would also be worthwhile to 
make a consistency check of the model in radiative decays, such as 
$\psi' \to \chi \gamma$, which are sensitive to the spatial structure 
of the $\psi'$ wave function beyond the $\sim 1/m_Q$ distances.

The implications of the color octet model for quarkonium hadroproduction 
were discussed at this Workshop by {\bf M. Beneke}. The main emphasis 
in this talk was on a rather crucial issue -- does factorization apply 
to quarkonium hadroproduction? On intuitive level, one expects 
soft gluon interactions between the $\bar{Q}QX$ system and the remnants of 
the hadrons to cancel at high $p_t$, up to corrections on the order of 
$(\Lambda_{QCD}/p_t)^n$. The total hadroproduction cross section is however 
dominated by the region of small $p_t$, and in this region factorization 
can hold only for sufficiently heavy quarks up to the terms of the order 
$(\Lambda/m_Q v^2)^k$. If $\Lambda$ in this estimate is $\Lambda \simeq 
\Lambda_{QCD}$, the ratio $(\Lambda/m_Q v^2) \simeq 1/3$ for \J, 
which makes factorization questionable. Moreover, one may expect 
that the parameter $\Lambda$ is actually proportional to the strength of 
the color field/density of partons produced in the collision, and this 
would make factorization in inclusive \J\ production even more questionable. 
The issue becomes really crucial in nucleus-nucleus collisions  
at high energies, where the number of produced partons which can interact 
with the $\bar{Q}QX$ system becomes very large.
Needless to say, it is very important to verify factorization in quarkonium 
production, for example, by a more detailed comparison of hadro- and 
photo-production of \J, including the polarized production to get rid 
of some of the uncertainties in the values of the color octet matrix 
elements.             

The 1987 discovery 
of $J/\psi$ suppression in nucleus-nucleus collisions 
by the NA38 Collaboration excited a lot of interest, to large extent 
because of the prediction \cite{MS} that this suppression would 
signal deconfinement, and would prove a highly non--trivial phase structure 
of QCD. The history of $J/\psi$ suppression, as summarized by {\bf H. Satz}  
at this Workshop, is a very dramatic one. Eventually, after a lot of 
controversy, the high-quality $pA$ data allowed to establish that, as 
was advocated by some theorists shortly after the NA38 results were 
announced,  
the observed in S-U collisions $J/\psi$ suppression can be explained 
in a very mundane way by what is often called the ``$J/\psi$ absorption". 
I put this expression in quotation marks, because, at a closer look, none 
of the words used to describe this phenomenon are correct: i) 
the fact that $\psi'/J/\psi$ ratio in $pA$ does not depend on atomic number 
at positive $x_F$ {\it proves} that what is ``absorbed" is not a 
physical $J/\psi$; 
ii) the $c\bar{c}$ pair cannot be really ``absorbed" by nucleons 
-- the only thing that can cause the observed $J/\psi$ depletion 
is a reduction of the probability that the pair will form a $J/\psi$, accompanied by 
an increased yield of open charm mesons\footnote{Unfortunately, since the 
\J\ yield is only a tiny fraction of the open charm one, this small increase 
in the production of open charm would be extremely difficult, if not 
impossible, to establish.}. 

I would like to stress that the high quality $pA$ data are absolutely 
necessary to understand the physical meaning of the effect observed 
in nuclear  
collisions. Fortunately, the $pA$ data become more and more accurate 
and extend to the kinematic region which was not explored before -- 
the new data from Fermilab E866 were presented at the Workshop by 
{\bf M. Leitch}. For the first time, the  
E866 Collaboration are able to measure \J's and $\psi'$'s 
in the negative $x_F$ window -- down to $x_F \simeq -0.15$. 
What we see through this window is exciting -- the $\psi'/J/\psi$ ratio, 
which was known to be independent of the atomic number, starts to 
drop toward negative $x_F$. I cannot refrain from saying that this 
is exactly what was predicted to 
happen (see e.g. Ref. \cite{BM} and Fig.2 in Ref. \cite{KS}) -- 
at negative $x_F$, the 
$\bar{c}c$ pair becomes slower, and starts to form inside the nuclear 
target. Since the $\psi'$ is much larger than the \J, it is absorbed 
with a larger probability, and the  $\psi'/J/\psi$ ratio drops. 
It now looks possible to extract the formation time of the pair 
directly from the data; a superficial first look suggests the 
proper formation time of $\tau \sim 0.3\ {\rm fm}$. One of the top 
priorities for the theorists now is to understand the detailed 
shape of the \J\ and $\psi'$ distributions in the entire range 
of $x_F$; we no longer have an excuse of insufficient data. 
The current state of affairs in explaining these distributions 
can hardly be considered satisfactory -- usually one has to invoke 
an eclectic mix of different mechanisms, such as shadowing, absorption, 
energy loss etc, the magnitude of each of which is poorly known. 

The \J\ production in proton--nucleus collisions and the role of 
the color-octet mechanism in $\bar{c}c$ absorption were discussed 
at this Workshop by {\bf C.-Y. Wong} and {\bf X.-F. Zhang}. 
This is clearly a very interesting problem which is central 
to the understanding of ``conventional" background to the 
\J\ suppression. 
In both talks, the absorption was considered as arising from 
the interaction of $\bar{c}c$ pair both in the color-octet and color-
singlet states. The incoherent mixture of these states was assumed 
in both talks, with the survival probability
\be 
S_A = f_1\ exp(-L \rho \sigma^1_{abs}) + f_8\ exp(-L \rho \sigma^8_{abs}),  
\ee
where $f_{1,8}$ are the fractions of \J\ coming from the color-singlet 
and color-octet mechanisms, and $\sigma^{1,8}_{abs}$ -- the corresponding 
absorption cross sections of the $\bar{c}c$ pair, $\rho$ -- mean 
nuclear density and $L$ - the mean path of the pair in nuclear matter. 
This assumption 
may be questioned since the color exchanges between the pair and the 
nuclear medium will rotate the color polarization vector of the pair 
as it traverses the nucleus. The mixture of the color-octet and color-singlet 
components in the $\bar{c}c$ wave functions will therefore become 
coherent, and the whole problem becomes a classical case for a 
coupled channel calculation.  
 
Let me come now to the CERN Pb-Pb results from the NA50 Collaboration. 
{\bf E. Scomparin} presented at this meeting intriguing new data on 
``intermediate mass enhancement" in the dilepton spectra. We have seen 
a rather dramatic, up to a factor of 3, (compared to the pA collisions) 
enhancement of the intermediate mass (below \J) dilepton yield 
in central Pb-Pb collisions. The origin of this enhancement is unknown; 
one of the explanations discussed by Scomparin was the 
enhancement of open charm, manifesting itself, through leptonic decays, 
in the enhancement of dilepton yield. If this were indeed true, the 
consequences of this phenomenon would be far--reaching -- we would have 
to accept  
an extremely strong (a factor of 3!) violation of factorization 
in the heavy quark 
production on nuclear targets. All of the existing phenomenology of hard 
processes in nuclear collisions would be ruined; of particular importance 
for the subject of this Workshop would be that the ``$\bar{c}c$ absorption" 
picture of \J\ suppression in pA and AB collisions would have to be 
abandoned, since all of the existing calculations rely on factorization 
in computing the initial yield of heavy quarks. 

However, I think we are still far from reaching these, rather dramatic,  
conclusions, since it is not yet clear that the origin of the 
intermediate mass enhancement is indeed the enhancement of the total 
yield of open charm. 
Other possibilities, including redistribution of heavy quarks in phase space 
to enhance the dilepton yield in the NA50 coverage \cite{LW}, secondary meson-meson 
interactions, and perhaps other novel phenomena have to be carefully 
considered before the enhancement of open charm is considered to be 
reliably established. 
Of course, a {\it direct} measurement of open charm yield would be 
extremely desirable to clarify the situation. In my opinion, this measurement 
should be considered  
as a high priority experiment for the future heavy ion program -- it 
would dramatically reduce the amount of uncertainty we have to face at 
present. 

The eagerly awaited data on charmonium production were presented at 
the Workshop by {\bf M. Gonin}. The data confirmed the presence of strong 
\J\ suppression in Pb-Pb collisions, going way beyond the expected 
on the basis of extrapolation of pA and S-U results. 
Also presented were the distributions of \J's and Drell-Yan pairs 
as a function of the produced transverse energy, which carry 
a significant amount of additional, as compared to $J/\psi/DY$ ratios, 
information. These data, not surprisingly, triggered the next round of 
hot discussions at the Workshop. {\bf A. Capella} suggested that 
the ``discontinuity" observed in the $J/\psi/DY$ ratio was only apparent, 
since it was less pronounced in the \J\ distribution in $E_T$ taken 
separately. The discussion which followed clearly demonstrated that 
the $E_T$ distributions of \J's and Drell-Yan pairs contain far more 
information than the ratios themselves. In particular, Drell-Yan pair 
$E_T$ distributions reflect the correspondence between the number of binary 
collisions (since the number of Drell-Yan pairs is determined by them) 
and the amount of produced transverse energy $E_T$. This correspondence 
is a very sensitive measure of the hadron production dynamics, and 
can be used to test the scaling of the number of produced hadrons 
with the number of participants (or binary collisions), which is 
significantly different for various models. In a calculation presented 
by Capella, the number of produced hadrons was evaluated as a sum 
of two terms -- the first was proportional to the number of participants 
and the second -- to the number of binary collisions. Since the number 
of binary collisions significantly increases in going from S-U to Pb-Pb, 
both the number and the density of the produced hadrons significantly grow, 
and it becomes possible to explain the {\it absolute magnitude} of the 
observed suppression. Of course, the model does not predict any 
discontinuities or structures in \J\ yield as a function of $E_T$, and 
experimental confirmation (or disproof) of the existence of such structure 
becomes quite important. It is also important to remember that even 
under the most dramatic assumptions about the \J\ suppression 
as a function of energy density, the fluctuations in the correspondence 
between $E_T$ and the impact parameter significantly smear out 
the discontinuity 
in the \J\ survival probability. Therefore, when looked at a higher resolution 
(more bins in $E_T$) the discontinuity in the observed \J\ distribution 
would be somewhat washed out, but would still possess a rather 
distinctive structure. Even if this structure were not present in the data, 
we still need to check if the models invoked to explain the magnitude of 
the suppression are consistent with {\it all} of the observations, 
including the Drell-Yan pair and \J\ $E_T$ distributions, minimum bias 
$E_T$ distributions and the correlation 
between the transverse $E_T$ and forward $E_{ZDC}$ energy.  

The possibility to explain the Pb-Pb data in the framework of conventional 
models was also discussed by {\bf R. Vogt}. 
The model with ``comover" absorption was considered, with  
the conclusion that this model was incapable of describing even the 
absolute magnitude of the observed suppression. Similar conclusion 
has also been reached in a different approach by {\bf C.-Y. Wong}. 
Clearly, we have to 
continue scrutinizing the existing models and confronting them with 
all of the available experimental data. 

One of the crucial inputs to the ``comover" calculations is the value of the 
\J\ absorption cross section in its interactions with mesons. 
Theoretical predictions for this quantity differ by almost three 
orders of magnitude 
(see e.g. Ref. \cite{KS1} and Ref. \cite{Bl}). A new calculation was presented 
at the Workshop by {\bf B. M\"uller}. It is based on the evaluation of 
the diagrams with $t-$ channel $D$ meson exchange, contributing to 
the exclusive processes of \J\ dissociation, e.g. $J/\psi + \rho \to D + 
\bar{D}$. The reasonable values of the corresponding coupling constants 
lead to the dissociation cross section which in the relevant energy 
region is on the order of $0.1\ {\rm mb}$. It had been noted however 
that the $D$ meson exchanged in the $t-$channel of the dissociation 
process is far off its mass shell. This has two implications: 
first, one has to consider heavier ($D^*,...$) exchanges, and second, 
one has to take account of the form factors. Little is known about 
the $t-$dependence of $D$ meson couplings, but a conservative guess 
leads to the reduction of the cross section by an order of magnitude. 
The resulting absorption cross section then becomes on the order of 
$0.01\ {\rm mb}$, which is a value fully consistent with the results of 
Ref. \cite{KS1}. It should be noted that in both approaches the smallness 
of the dissociation cross section stems from the large mass of the 
charm quark mass. If the cross section is indeed so small, the rate 
of \J\ dissociation in a hadron gas is negligible. 
It would be important to measure the \J\ dissociation cross section 
at small energy directly, either in the inverse kinematics experiment, or in 
the experiment utilizing Fermilab antiproton accumulator and nuclear 
jet target to 
produce low-momentum ($\sim 4\ \rm{GeV/c}$) \J's in the process 
$\bar{p}A \to J/\psi + (A-1)$.     
  
Another effect potentially contributing to the suppression of \J\ 
production -- depletion of the gluons in nuclear medium -- 
was discussed by {\bf R. Hwa}. This approach is based on the observation  
that at high energies, because of the Lorentz factor,   
the time during which the nucleon traverses the nucleus is shorter  
than it takes   
for a signal to propagate through the nucleon's   
transverse size. This implies that one can distinguish   
between the interactions of (anti)quarks and gluons from the incident   
nucleon. Because of the larger color charge, the gluons are expected   
to interact stronger than the quarks inside the nucleus -- the 
nucleus therefore can 
act as a gluon filter!   
Since the Drell-Yan pairs are   
produced (in the leading order in $\alpha_s$) by the quark-antiquark fusion,   
and the heavy quarks by the gluon-gluon fusion, one can try to  
reconcile the absence   
of initial state effects in Drell-Yan pair production with the strong  
suppression        
observed for the \J\,even though the Drell-Yan data still do impose an   
important constraint on the model.     
 
Besides \J\ suppression, this mechanism should also cause suppression of  
the open charm production in $pA$ and $AB$ collisions.     
Even though the current data do not seem to show such suppression,  
 more data, particularly on correlated $\bar{D}D$  
production, are needed to clarify the issue. Nevertheless,  
before such data become available, one  
can also argue in favor of {\it universality} of quark and gluon  
depletion in nuclear matter at small $x$ (high energies and central  
region), which implies that the initial-state gluon absorption  
(or energy loss) is unlikely to be  
{\it the} mechanism responsible for the observed \J\ suppression.     
Indeed, the virtuality ordering in the QCD DGLAP evolution  
means that at small $x$, the heavy quarks and Drell-Yan pairs  
are generated  
{\it at the very end} of the parton ladder; the evolution at the  
preceding stages of the parton cascade is identical in both cases.  
Moreover, the gluons fusing to form a heavy quark pair, or quarks  
and antiquarks annihilating into a Drell-Yan pair, have a large  
virtuality and, at small $x$, small momentum -- therefore, they almost do not  
propagate inside the nucleus!  
The situation will change, however, if we move out of the central region,  
since either $x_p$ or $x_t$ will then become large, involving  
the valence partons in the production process -- in this case the nucleus 
can indeed act as a gluon filter, suppressing gluon--induced processes 
stronger than the processes induced by the valence quarks. 
This is a very interesting topic to study! 

The formation time effects in the production of Drell--Yan pairs 
were addressed at the Workshop by {\bf J. Kapusta}. The motivation 
for this study is a well--known discrepancy between the Drell--Yan data and  
the predictions of probabilistic cascades. In a typical 
cascade calculation, the nucleon loses energy immediately after an 
inelastic collision with a nucleon inside the nuclear target. Since 
at CERN SPS and Fermilab fixed target experiments we are still in the 
energy range where the Drell--Yan production cross section is a steep 
function of the incident nucleon's momentum, this initial state energy 
loss will lead to a strong suppression of the Drell--Yan production. 
This strong suppression in the Drell--Yan yields, however, is not seen  
experimentally, and this is a major problem for the probabilistic cascades.
Even though this problem is well--known to insiders, so far 
this failure was not clearly acknowledged and documented by 
a direct comparison of a cascade calculation to the data. 
This comparison was shown at the Workshop by Kapusta, who demonstrated 
that a cascade calculation with energy loss 
under-predicts the Drell--Yan yields by a factor 
of up to $\sim 3$. It has also been stressed that this problem 
is a direct consequence of quantum mechanics, and to solve it within a 
cascade calculation, one has 
to simulate quantum--mechanical effects. This has been done by introducing 
the formation time, during which the incident nucleon does not ``know" yet 
whether the energy has been lost.  
\J\ suppression 
in the framework of a different numerical cascade model was presented at the 
Workshop by {\bf J. Geiss}. 

The difficulties of conventional approaches motivated several theorists 
to propose that the \J\ suppression observed in Pb-Pb collisions is, 
after all, 
the signal of deconfinement, as was originally proposed \cite{MS}.   
An interesting realization of the deconfinement scenario was presented  
at this Workshop  
by {\bf E. Shuryak}; 
in this approach, the produced deconfined   
phase reaches its ``softest point'' at some centrality in Pb-Pb collisions.  
This leads to a very long lifetime of the produced plasma, which can  
therefore effectively dissociate the produced \J's.   
A distinctive feature of this approach is that the \J\ absorption  
is maximal at some value of centrality, corresponding to the ``softest point''  
of the equation of state of the produced deconfined phase;  
once the centrality increases further, the \J\ survival probability increases again. 
However, also in this approach, the sharp discontinuity is difficult to explain, and we  
are still left with the ``jump puzzle''. 
 
An approach aiming at the explanation of the observed structure in the \J\ survival 
probability was presented in the talks of {\bf M. Nardi} and {\mbox {\bf H. Satz}}. 
In this approach, one considers the strings produced in 
nucleon-nucleon interactions in the transverse plane. Each of the strings 
is characterized by some transverse size, and when the density of strings becomes 
large enough, they ``percolate", forming connected clusters. 
The assumption then is that this percolation corresponds 
to the formation of deconfined matter in which the \J's are dissociated. 
While this is an interesting and attractive picture, a number of questions have to be 
answered. The first, and most natural, question concerns the nature of ``strings": 
do they represent inherently soft interactions, or perhaps mini-jets? (In the latter case 
the transverse size of the ``string" would be determined by the transverse momentum 
of the produced mini-jet, $r_{\perp} \sim 1/k_{\perp}$.)  
It is also not clear why the number 
of produced ``strings" is proportional to the number
of collisions, while the number of produced 
secondaries looks to be proportional to the number of 
``wounded" nucleons. One can argue that 
the strings can fuse, but  
this would imply a very strong interaction
among them, which is not taken into
account in the percolation picture. On the other hand, 
in the independent string picture, it is
unclear how the strings can fuse. 
Despite of these questions, the percolation provides an attractive 
and potentially enlightening framework for the description of deconfinement 
in heavy ion collisions, which has to be investigated in more detail. 

Most of the issues described above were the subject of hot debates 
during the round table discussion led by {\bf M. Gyulassy}. 

\section{The future of quarkonium}
  
The future of quarkonium physics was addressed at the Workshop by 
{\bf Y. Akiba, T. LeCompte, C. Louren\c{c}o} and {\bf M. Rosati.} 
{\bf Y. Akiba and \mbox{M. Rosati}} discussed the capabilities of the PHENIX detector 
at RHIC in doing quarkonium physics. With the possibility to reconstruct 
approximately half a million of \J's 
in the dimuon decay mode, and fifty thousand in the 
dielectron decay during one year of running, the future of quarkonium physics at RHIC 
looks very bright. The PHENIX program will be nicely supplemented by the 
quarkonium program of STAR, as was discussed by {\bf T. LeCompte}. 
STAR will be able to detect \J's at high transverse momentum in the dielectron 
decay mode -- for the first time in the history of \J\ suppression in nucleus-nucleus 
collisions, we are going to have two independent experiments 
producing the data which can actually be compared and checked against each other! 
At RHIC we will finally gain access to the study of $\Upsilon$ suppression in nuclear 
collisions; 
the expected statistics -- about one thousand events per year in the dimuon mode, 
as given by {\bf M. Rosati} for the PHENIX acceptance -- is more 
modest here, but it would be very interesting to glean at least some information about 
the flavor dependence of quarkonium suppression. Since the $\Upsilon$ is very tightly 
bound and has the binding energy more than twice larger than that of the \J, 
it would probe the presence of very hard gluons in the medium; in the thermal picture, 
the dissociation of the $\Upsilon$ would require extremely high temperatures. 
As was noted by {\bf Y. Akiba}, quarkonium production in polarized $pp$ collisions 
will help to understand better the production mechanism. {\bf C. Louren\c{c}o} 
presented a summary of the experimental results on \J\ production, and outlined the
issues of potentially large importance for quarkonium physics at colliders -- 
notably, the necessity to be able to separate direct charmonium production from 
that resulting from $B$ decays.   

Of course, the expected rates of \J\ production at RHIC depend crucially on the 
magnitude of the suppression. In the most optimistic (for theorists) scenario of strong 
suppression, 
one may even wonder if there will be enough \J's to detect. I do not think, however, that 
the experimentalists should worry about this -- even if {\it all} of the \J's 
inside the quark--gluon plasma 
are absorbed, we have to remember that, as evidentiated by the Tevatron results, 
at high $p_t$ most of the \J's are the products of the fragmentation 
of high--$p_t$ gluons. Since any number of rescatterings inside the plasma will not 
affect the probability of the gluon fragmentation to a \J, there would be a component 
in \J\ production that will survive even in the most central collisions. However the  
re-scatterings inside the system can change the  $p_t$ distribution of the gluons, 
and therefore 
the resulting $p_t$ distribution of the produced \J's, shifting them to smaller momenta. 
Therefore, at high $p_t$ \J\ production at collider energies may be sensitive 
to the gluon 
energy loss. The corresponding theory and phenomenology still have to 
be developed.   

One should also remember that at collider energies, we will enter 
the gluon shadowing domain in \J\ production, and the shadowing will  
become important even for the production at central rapidity. To separate the 
shadowing effects, 
we would certainly need to have $pA$ data as well. And, to have a proper reference 
point, the $pp$ data on quarkonium production in both unpolarized and polarized 
collisions are indispensable.
\vskip0.3cm
The Workshop in Seattle has shown that the physics of heavy quarkonium 
continues to develop at a fast pace; moreover, we have every reason to believe that 
the next year, with RHIC turning on, will mark the beginning of the new exciting era 
in this field.

\section*{Acknowledgments}
 
I would like to thank X.-N. Wang, B. Jacak, J. Kapusta and C.-Y. Wong 
for making this Workshop extremely interesting and productive. 
I thank also the Institute of Nuclear Theory for support, 
and its staff for providing the most 
comfortable environment for the Workshop.

\end{document}